\documentclass[pra,twocolumn,showpacs,superscriptaddress]{revtex4}
\usepackage{graphicx}
\newcommand{\nn}{\nonumber}

\begin{document}

\title{Axisymmetric versus Non-axisymmetric Vortices 
                                 in Spinor Bose-Einstein Condensates}

\author{T. Mizushima}
\email{mizushima@mp.okayama-u.ac.jp}
\affiliation{Department of Physics, Okayama University,
             Okayama 700-8530, Japan}
\author{K. Machida}
\affiliation{Department of Physics, Okayama University,
             Okayama 700-8530, Japan}
\author{T. Kita}
\affiliation{Division of Physics, Hokkaido University,
             Sapporo 060-0810, Japan}
\date{\today}

%>>>########################### ABSTRACT ############################## 
\begin{abstract}
The structure and stability of various vortices 
in $F\!=\!1$ spinor Bose-Einstein condensates 
are investigated by solving the extended Gross-Pitaevskii equation under rotation.
We perform an extensive search for stable vortices, considering both axisymmetric and non-axisymmetric vortices 
and covering a wide range of ferromagnetic and antiferromagnetic interactions.
The topological defect called Mermin-Ho (Anderson-Toulouse) vortex 
is shown to be stable for ferromagnetic case.
The phase diagram is established in a plane of external rotation $\Omega$ vs 
total magnetization $M$ by comparing the free energies of possible vortices. 
It is shown that there are qualitative differences between axisymmetric 
and non-axisymmetric vortices which are manifested in the $\Omega$- and $M$-dependences.
\end{abstract}
%<<<########################### ABSTRACT ##############################


\maketitle

%>>>########################## INTRODUCTION ##########################
\section{Introduction}

The experimental achievement of Bose-Einstein condensation (BEC) 
in the trapped atomic clouds\cite{cornell,bradley,davis} has opened up 
a novel field to investigate fundamental problems such as the relationship 
between the superfluidity and BEC. 
Owing to recent advances of experimental techniques, 
several groups have succeeded in creating quantized vortices with various procedures 
in the magnetically trapped BEC\cite{matthews,madison,abo,haljan,hodby,leanhardt}, 
where the condensate is described by a scalar order parameter. 
Furthermore, the atomic gases with the hyperfine spin $F=1$ called ``spinor BEC'' 
have been Bose-condensed via optical methods\cite{stenger,barrett} 
which can keep the atomic ``spin'' states degenerate and active.
As shown recently by Klausen {\it et al}.\cite{klausen}, 
the spin-dependent interaction of two $^{87}$Rb atoms is ferromagnetic. 
Thus, we now have spinor BEC's with both antiferromagnetic ($^{23}$Na)\cite{stenger}
and ferromagnetic interactions.

Such scalar and spinor BEC's are analogous to superfluid $^3$He and $^4$He. 
These superfluid Heliums, however, have rather strong interactions.  
Indeed, the condensate fraction in superfluid $^4$He is only 10$\%$ of the total. 
By contrast, BEC of the atomic gases have advantages 
for both theoretical and experimental treatments
due to their weak interactions; 
here almost all the atoms are able to be Bose-condensed.
It is possible to directly observe dynamical behaviors of the condensate
with optical methods,
providing us an opportunity to quantitatively investigate
the new quantum fluid\cite{dalfovo}.
 
The standard Hamiltonian for the spinor BEC  
have been introduced by Ohmi and Machida\cite{ohmi},
and Ho\cite{ho}, who pointed out the richness of the exotic topological defects.
Topological structures, such as
skyrmion, meron, Mermin-Ho (Anderson-Toulouse) texture, and monopole,
play an important role in various fields of physics.
They provide a common framework to connect diverse field,
thereby enhancing mutual understanding\cite{jayaraman, leggett}.
Specially, since $F\! =\! 1$ ferromagnetic BEC can be described by order parameters
similar to the superfluid $^3$He-A, the coreless Mermin-Ho vortex 
may be favored in the ferromagnetic BEC\cite{mermin, anderson}.

Similar topological structures, called skyrmion in general, have been proposed 
in the spinor BEC. 
Al Khawaja and Stoof \cite{stoof} studied a skyrmion in the 
$F\! =\! 1$ ferromagnetic BEC and concluded that it
is not a thermodynamically stable object without rotation.
By considering the effect of the external rotation, however, 
we have shown recently\cite{mizushima} that 
this topological defect can be stable.
Ingenious proposals have been made\cite{marzlin,anglin,kurihara,martikainen} on 
how to create it and detect it.
Yip\cite{yip} has performed a systematic study on vortex structures
and presented several 
axisymmetric and non-axisymmetric vortices for $F\! =\! 1$ antiferromagnetic BEC.
Recently, Isoshima {\it et al}.\cite{isoshima1,isoshima2}
have carried out an extensive study of axisymmetric vortices to provide a
vortex phase diagram in a plane of the rotation and the magnetization 
for both the antiferromagnetic and ferromagnetic cases.

In this paper we examine the stability of various vortices  
for both the $F\! =\! 1$ antiferromagnetic and ferromagnetic BEC
trapped in a two-dimensional harmonic potential. 
We have removed the previously imposed restriction in the axisymmetric case that
winding numbers are less than or equal to unity.
The continuous vortices such as the Mermin-Ho vortex will also be shown 
to be favored over the singular ones\cite{isoshima1,isoshima2} 
and the other non-axisymmetric ones\cite{yip}.
We demonstrate the stability of such vortices and discuss differences
between the axisymmetric and non-axisymmetric configurations.
We also determine the vortex phase diagrams for the antiferromagnetic and ferromagnetic cases.
By comparing the relative free energies of the possible vortex configurations 
and the phase-separated state which may occur in the ferromagnetic situation, 
the validity of assuming uniformity along $z$ direction
will be checked.

This paper is organized as follows.
In Sec. II, we first present the extended Gross-Pitaevskii equation for the spinor BEC, 
and then explain the numerical procedure to find local minima of the energy functional.
Section III enumerates possible vortices 
for axisymmetric and non-axisymmetric cases, 
and then discuss the stability of the Mermin-Ho vortex.
Section IV presents the phase diagram for the ferromagnetic and antiferromagnetic cases in the plane of 
external rotation $\Omega$ vs total magnetization $M$ obtained 
by comparing the free energies.
Here, the qualitative differences between the axisymmetric and non-axisymmetric vortices 
are discussed by showing the $\Omega$- and $M$-dependence. 
The final section is devoted to conclusions and discussions.

%%%---<<< Revised in 2002.07.24 

%>>>########################### FORMULATION ##########################
\section{Theoretical Formulation}

%*************** Extended Gross-Pitaevskii equation ***********
\subsection{Extended Gross-Pitaevskii equation}

We consider Bose condensed $F\!=\!1$ spinor BEC's with internal degrees of freedom
for both ferromagnetic and antiferromagnetic cases.
Here the order parameters are characterized by the hyperfine sublevels $m_F = 1, 0, -1$.
We start with the standard Hamiltonian by Ohmi and Machida\cite{ohmi}, and
 Ho\cite{ho}: 
%==================== Hamiltonian =======================================
\begin{eqnarray}
\hat{{\mathcal H}}_{rot} 
     & = & \hat{{\mathcal H}} - \int\! d{\bf r} {\bf \Omega} \cdot 
                                \sum_{j} \Psi_j^{\dagger}
                                ({\bf r} \times {\bf p})\Psi_j \nn \\
\hat{{\mathcal H}} 
     & = & \int d{\bf r} 
           \left[
                 \sum_{ij}\Psi_i^{\dagger}
                   \left\{ h({\bf r}) - \mu_{i} \right\}
                   \Psi_j \delta_{ij} \right. \nn \\
     &   &       + \frac{g_{\rm n}}{2} \sum_{ij}
                   \Psi_i^{\dagger} \Psi_j^{\dagger} 
                   \Psi_j \Psi_i \nn \\
     &   &       + \left. 
                   \frac{g_{\rm s}}{2} \sum_{\alpha}
                   \sum_{ijkl} \Psi_i^{\dagger}\Psi_j^{\dagger}
                   (\hat{F}_{\alpha})_{ik}(\hat{F}_{\alpha})_{jl}
                   \Psi_k \Psi_l
           \right].
\label{eq:hamiltonian}
\end{eqnarray}
%==================== Hamiltonian =======================================/
Here
\begin{eqnarray} 
   h({\bf r}) = - \frac{\hbar^2 \nabla^2}{2m} + V({\bf r})
\end{eqnarray}
is one-body Hamiltonian.
The quantity $V({\bf r})\!=\!\frac{1}{2} m (2\pi\nu_r)^2 (x^2 + y^2)$ 
 is the external confinement potential
 such as an optical potential.
The scattering lengths $a_{0}$ and $a_{2}$ characterize collisions between
 atoms through the total spin 0 and 2 channels, respectively,
 $g_{\rm n} = \frac{4 \pi \hbar^2}{m} \cdot \frac{a_0 + 2a_2}{3}$ is 
 interaction strength through the ``density'' channel,
 and $g_{\rm s} = \frac{4 \pi \hbar^2}{m} \cdot \frac{a_2 -  a_0}{3}$ is 
 that through the ``spin'' channel.
The subscripts $\alpha = (x,y,z)$ and $i,j,k,l = (0, \pm1)$
 correspond to the above three species.
The chemical potentials for the three components
 $\mu_i$ $(i = 0, \pm1)$ 
 satisfy $\mu_1 - \mu_{0} = \mu_{0} - \mu_{-1}$.
We introduce $\mu = \mu_{0}$ and $\mu' = \mu_{1} - \mu_{0}$.
The angular momentum operators $\hat{F_{\alpha}} (\alpha = x, y, z)$  
can be  expressed in matrices as 
%================ Angular Momentum oparators ====================
\begin{eqnarray}
 F_x = \frac{1}{\sqrt{2}}
       \left( 
              \begin{array}{ccc}
              0 & 1 & 0\\
              1 & 0 & 1\\
              0 & 1 & 0
              \end{array}
       \right), \nn \\
 F_y = \frac{i}{\sqrt{2}}
       \left(
              \begin{array}{ccc}
              0 & -1 &  0\\
              1 &  0 & -1\\
              0 &  1 &  0
              \end{array}
       \right), \nn \\
 F_z =
       \left( \;
              \begin{array}{ccc}
              1 &  0 &  0\\
              0 &  0 &  0\\
              0 &  0 & -1
              \end{array}
       \right).
\end{eqnarray}
%================ Angular Momentum oparators ====================/

Following the standard procedure, 
the extended Gross-Pitaevskii (GP) equation in rotation frame is obtained as 
%================ GP equation ====================
\begin{eqnarray}
    [
      \{
          h({\bf r}) - \mu_i 
      & + & g_{\rm n} \sum_l |\psi_l|^2 
            \} \delta_{ij} \nn \\ 
      & & + g_{\rm s} \sum_{\alpha}\sum_{kl} 
              \{
                (F_{\alpha})_{ij} 
                (F_{\alpha})_{kl} \psi_k^{\ast} \psi_l
              \}  \nn \\ 
      & & - i \hbar {\bf \Omega} \cdot \nabla \times {\bf r} \delta_{ij}
    ] \psi_j 
   = 0.
\label{eq:gp}
\end{eqnarray}
%================ GP equation ====================/
These coupled equations for the $j$-th condensate wave function
$\psi_j=\langle \Psi_j \rangle$  $(j = 0, \pm1)$ are used to calculate
various properties of  vortices in the following.
%Here we assume uniformity along $z$ direction and take the external rotation  
%as ${\bf \Omega} \!=\! \Omega \hat{{\mbox{\boldmath $z$}}}$.
Here we take the external rotation as ${\bf \Omega} \!=\! \Omega \hat{{\mbox{\boldmath $z$}}}$ 
and assume uniformity along z direction.

%********************* Numerical procedure **************
\subsection{Numerical procedure}

The stationary states of the extended GP equation are defined as 
local minima of the energy functional 
%================ Energy Functional =============================
\begin{eqnarray}
 E [\psi_i, \psi_i^{\ast} ] 
   = \int d {\bf r} \left[ \sum_i E_i ({\bf r}) + E_s ({\bf r}) \right]
     - \mu' M
     - {\bf \Omega} \cdot {\bf L},
\label{eq:eneFn}
\end{eqnarray}
where $E_i$ and $E_s$ are defined by 
\begin{eqnarray} 
 E_i ({\bf r})   
   & = & \psi_i^{\ast} \left\{ h ({\bf r}) - \mu
                             + \frac{g_n}{2} \sum_k | \psi_k |^2 
                     \right\} \psi_i, \\ 
 E_s ({\bf r}) 
   & = & \frac{g_s}{2} \sum_{\alpha}      
                 \left\{ \sum_{k,l} 
                      ( \psi_k^{\ast} ( \hat{F}_{\alpha} )_{kl} \psi_l ) 
                 \right\}^2, 
\label{eq:each_eneFn}
\end{eqnarray}
%================ Energy Functional =============================/
\noindent
${\bf \Omega} \cdot {\bf L}$ denotes
\begin{eqnarray}
 {\bf \Omega} \cdot {\bf L} 
     = - i \hbar \Omega
           \int d{\bf r} \sum_i
                  \psi_i^{\ast} \left(x \frac{\partial}{\partial y} 
                    - y \frac{\partial}{\partial x} \right) \psi_i,
\end{eqnarray}
and $ M \!=\! \int d{\bf r} \sum_i i | \psi_i ({\bf r}) |^2$ is the total magnetization.

The numerical algorithm used to minimize the energy functional 
 in scalar BEC\cite{castin, garcia, aftalion, feder}
 can be extended to the present system.
Following this procedure, the initial $\psi_j$ given randomly
are modified using the local gradient of the energy functional
$E [\psi_i, \psi_i^{\ast} ]$ as
\begin{eqnarray}
    \psi_j ( \tau + \Delta \tau ) 
            = \psi_j ( \tau )
              - \frac{\delta E [\psi_i, \psi_i^{\ast}]}
                     {\delta \psi_j^{\ast}} 
                \Delta \tau.
\label{eq:ev}
\end{eqnarray}
This equation means that 
the order parameters $\psi_j$, parameterized by a `fictitious time' $\tau$,
roll along the slope of the energy functional.
Equation \ (\ref{eq:ev}) is rewritten as
\begin{eqnarray}
 - \hbar \partial_{\tau} \psi_j (\tau) 
   & = &\left[ 
              \left\{  
                      h ({\bf r}) 
                      - ( \mu(\tau) + \mu' j )
                      + g_n \sum_k | \psi_k |^2
              \right\} 
              \delta_{jk} \right. \nn \\ 
  &    &    + \frac{g_s}{2} \sum_{\alpha}
              \left\{ \sum_{l,p} 
                      ( \psi_l^{\ast} 
                        ( \hat{F}_{\alpha} )_{lp} 
                        \psi_p 
                      ) 
                      ( \hat{F}_{\alpha} )_{jk} 
              \right\} \nn \\
  &   & \left.  
            + i \hbar \Omega
              \left(x \frac{\partial}{\partial y}
                    - y \frac{\partial}{\partial x} 
              \right) \delta_{jk}
       \right] 
       \psi_k (\tau). 
\label{eq:imagGP}
\end{eqnarray}
This is the Gross-Pitaevskii equation for imaginary times $\tau=it$.
In each time step, $\mu(\tau)$ is adjusted to preserve the total number of particles in the system
\begin{eqnarray}
 N = \sum_j \int d{\bf r} | \psi_j ({\bf r})|^2.
\end{eqnarray}
For $\tau \rightarrow \infty$, 
 $\psi_j$ converges to the stationary state, 
 corresponding to one of the local minima of the energy functional \ (\ref{eq:eneFn}). 
For $\tau = \infty$ $\psi_j$ satisfies 
\begin{eqnarray}
 \left.
       \frac{\delta E [\psi_i, \psi_i^{\ast}]}
            {\delta \psi_j^{\ast}}
 \right |_{\tau \rightarrow \infty}
 = 0,
\end{eqnarray}
and Eq.\ (\ref{eq:imagGP}) becomes equivalent  
to the extended GP equation \ (\ref{eq:gp}).

We take the initial state of each component 
as 
\begin{eqnarray}
 \psi_j ({\bf r}, \tau=0) = \sqrt{n_{{\rm TF}}({\bf r})} \; \eta_j \exp[i S_j({\bf r})],
\end{eqnarray}
where $n_{TF}({\bf r})$ is the density profile within 
the Thomas-Fermi (TF) approximation: 
\begin{eqnarray}
n_{{\rm TF}}({\bf r}) = 
	\left\{ 
               \begin{array}{lc}
               	 \frac{\mu_{{\rm TF}} - V}{g_n}     & \mbox{for $g_s > 0$} \\
                 \frac{\mu_{{\rm TF}} - V}{g_n+g_s} & \mbox{for $g_s < 0$}
               \end{array}
        \right. ,
\end{eqnarray}
and $\eta_j$ represents the ratio of each component.
The phase is given by
\begin{eqnarray}
 S_j({\bf r}) = \sum_k w_j^{(k)} \theta_j^{(k)} + \alpha_j,
\end{eqnarray}
$w_j^{(k)}$ is the winding number of the $j$-th condensate, 
$\theta_j^{(k)}$ is the polar angle of the coordinate $(x^{(k)}, y^{(k)})$ 
whose origin is located at the $k$-th vortex core, 
and $\alpha_j$ is relative phase 
between the three components.

It is convenient to describe the condensates in terms of 
the three components $\psi_{\alpha}(\alpha=x,y,z)$ where the quantization axis 
is taken along the $\alpha$ direction:
\begin{eqnarray}
 \left( \begin{array}{c}
           \psi_{x} ({\bf r}) \\
           \psi_{y} ({\bf r}) \\
           \psi_{z} ({\bf r})
        \end{array}
 \right) 
        = \frac{1}{\sqrt{2}}
          \left( \begin{array}{ccc}
                    -1 &    0     &  1\\
                    -i &    0     & -i\\
                     0 & \sqrt{2} &  0
                 \end{array}
          \right)
 \left( \begin{array}{c}
           \psi_{1}  ({\bf r}) \\
           \psi_{0}  ({\bf r}) \\
           \psi_{-1} ({\bf r}).
        \end{array}
 \right) 
\end{eqnarray}
We then define a couple of real vectors as
\begin{eqnarray}
 \mbox{\boldmath $m$} = (m_x, m_y, m_z) = {\rm Re}(\psi_{x}, \psi_{y}, \psi_{z}), \\
 \mbox{\boldmath $n$} = (n_x, n_y, n_z) = {\rm Im}(\psi_{x}, \psi_{y}, \psi_{z}).
\end{eqnarray}
The $l$-vector, which points the direction of the local magnetization, is defined as 
$\mbox{\boldmath $l$}=\mbox{\boldmath $m$}\times\mbox{\boldmath $n$}$.
The corresponding unit vector is denoted by $\hat{\mbox{\boldmath $l$}}$.

%**************** Calculated system *************
\subsection{Calculated system}

The actual calculations are carried out by discretizing the two-dimensional
space into 51$\times$51 mesh. 
We have performed extensive search to find stable vortices, starting with various 
vortex configurations, covering a wide range of the ferromagnetic 
and the antiferromagnetic interaction
strength, $g_s/g_n= -0.2 \sim 0.2$, and examining various axisymmetric and 
non-axisymmetric vortices. See Ref.\cite{isoshima1,isoshima2} for 
the classification of possible vortices in the axisymmetric case.
We use the following parameters: 
the mass of a $^{87}$Rb atom $m$=1.44$\times 10^{-25}$kg, 
the trapping frequency $\nu_r$=200Hz, 
and the particle number per unit length along the $z$ axis $n_z$=2.0$\times 10^3 / \mu$m.
The results displayed here are for $g_s/g_n=-0.02$ (ferromagnetic case) 
and $g_s/g_n=0.02$ (antiferromagnetic case).
The external rotation frequency $\Omega$ is normalized by the harmonic trap frequency.

%###################### VORTEX STRUCTURE #############################
\section{vortex structure} 

The vortex configurations are characterized by the combination 
of the winding number $w_j$ of $\psi_j$ ($j\!=\!0$, $\pm 1$)
denoted by $\langle w_1, w_0, w_{-1} \rangle$,
where $w_j$ denotes the phase change by $2 \pi w_j$
when the wave function goes around the phase singularity.

The spin term (\ref{eq:each_eneFn}) of the total energy is rewritten as 
\begin{eqnarray}
  E_s ({\bf r})
          =  \frac{g_s}{2}
             \left[    n^2 ({\bf r}) 
                     - \left|
                         2 \psi_1 ({\bf r}) \psi_{-1} ({\bf r})
                         - \psi_0^2 ({\bf r})
                       \right|^2
             \right],
\label{eq:Es}
\end{eqnarray}
where $n({\bf r})=\sum_j |\psi_j({\bf r})|^2$ is the total density. 
The spin texture in the ground state without rotation 
is determined by this energy. 
By minimizing Eq.(\ref{eq:Es}), 
the relative phases $\alpha_j$ are shown to satisfy
\begin{eqnarray}
 2 \alpha_0 = \alpha_1 + \alpha_{-1} + n \pi,
\end{eqnarray}
where $n$ is an integer, and the odd (even) $n$ corresponds to 
the antiferromagnetic (ferromagnetic) situation\cite{isoshima1}.

%**************** Axisymmetric vortex **************
\subsection{Axisymmetric vortex}

It also follows from Eq.\ (\ref{eq:Es}) that $\langle w_1, w_0, w_{-1} \rangle$
of axisymmetric vortices satisfies
\begin{eqnarray}
2w_0 = w_1 + w_{-1}.
\end{eqnarray}
Thus, possible candidates for the stable state are the non-vortex state $\langle 0, 0, 0 \rangle$ 
and the vortex configurations:
$\langle 1, 0, -1 \rangle$, $\langle 1, \times, 0 \rangle$, $\langle 0, \times, 1 \rangle$,
and $\langle 1, 1, 1 \rangle$, which exhaust all the combinations of the winding numbers
less than or equal to unity.
The thermodynamic stability of these vortices are demonstrated in Ref.\cite{isoshima2}.
Here we concentrate on the possibility of combinations with higher winding numbers, 
which will be shown to be stable only in the ferromagnetic case.

Figure 1 displays the density and $l$-vector profiles of 
a new continuous vortex $\langle 0,1,2 \rangle$ found stable for the ferromagnetic interaction ($g_s\!<\!0$).
It is seen that $\psi_1$ with zero winding number $w_1 = 0$
occupies the central region of the harmonic trap
and $\psi_{-1}$ with the higher winding number $w_{-1} \!=\! 2$ 
fill in the circumference region. 
The intermediate region is occupied by $\psi_{0}$ component 
which has a singularity $w_0=1$ at the center of the trap. 
The resulting total density is non-singular 
and have a smooth spatial variation described by a Gaussian form
except for the outermost region. 
This vortex 
%has the winding number combination $\langle 0, 1, 2 \rangle$ in our notation and 
is equivalent to the topological structure called 
the Mermin-Ho vortex in superfluid $^3$He\cite{mermin} 
and Skyrmion in general\cite{stoof}. 

%===================== <012> ======================
\begin{figure}[b]
    \includegraphics[width=7cm]{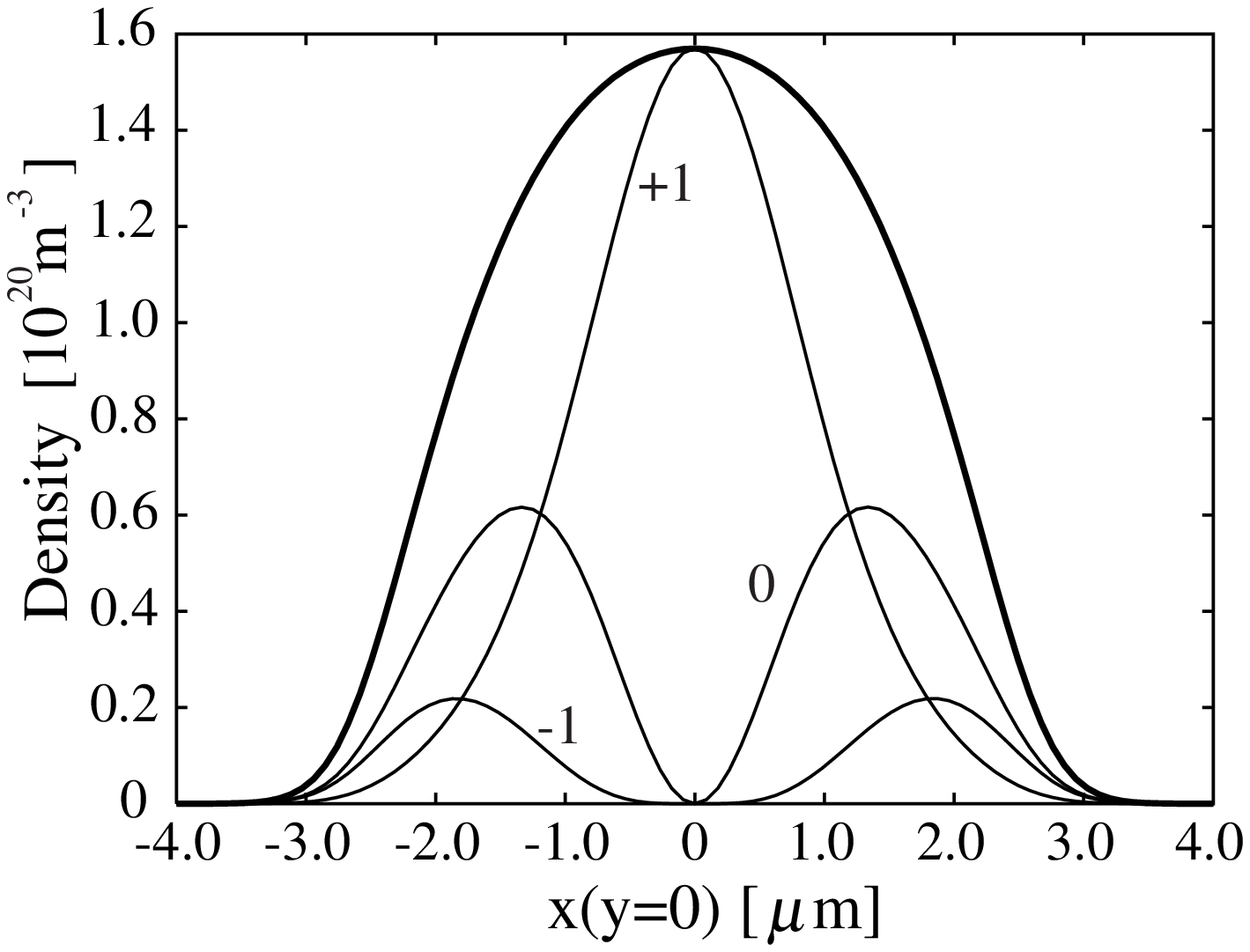}\\
    (a)	
 \begin{tabular}{cc} 
    \includegraphics[width=4.5cm]{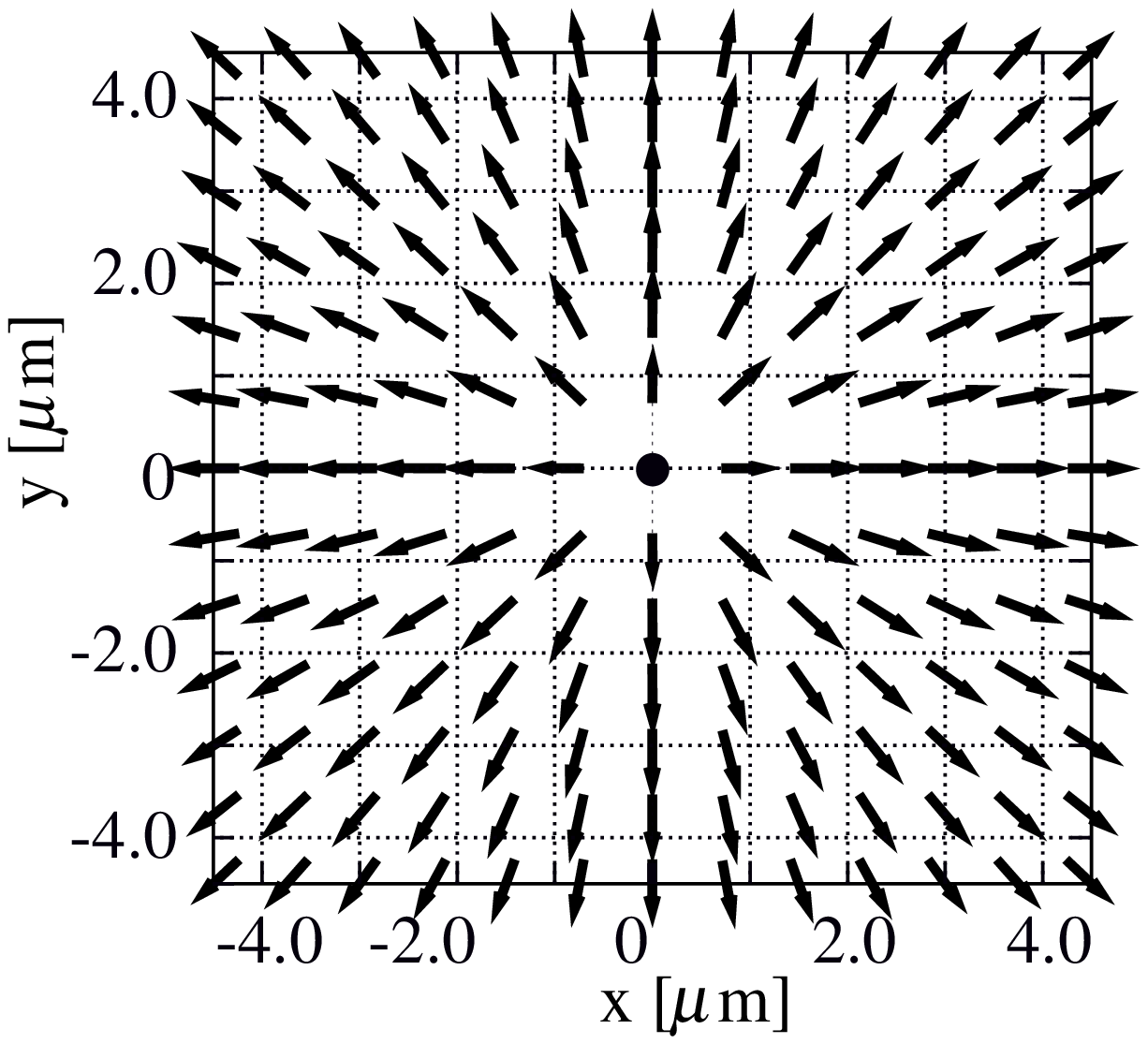}
    &
    \includegraphics[width=4cm]{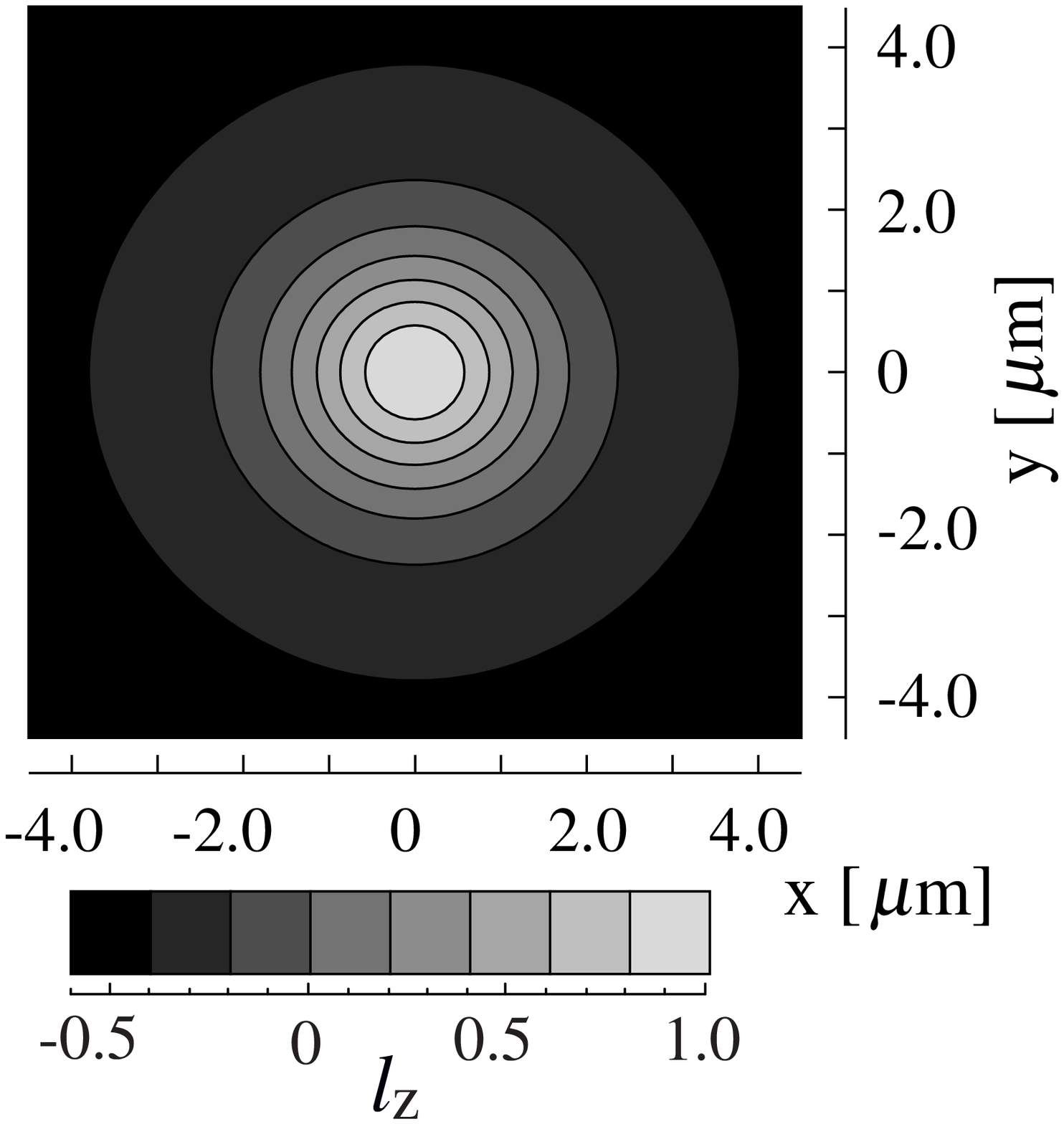}\\
    (b)   
    &               
    (c)             
  \end{tabular}     
\caption{\label{fig:012dns}
Properties of the $\langle 0,1,2\rangle$ vortex 
at $\Omega=0.35$ and $M/N\!=\!0.21$. 
(a) Density Profile; (b) $l_x$ and $l_y$; (c) density map of $l_z$.
The bold line in (a) denotes the total density $n({\bf r})$
and the thin lines show the density of each component $|\psi_j|^2$.
}
\end{figure}
%===================== <012> ======================/

The axisymmetric $\langle 0, 1, 2 \rangle$ vortex may be parametrized as
\begin{eqnarray}
 \left( \begin{array}{c}
           \psi_{1}  ({\bf r}) \\
           \psi_{0}  ({\bf r}) \\
           \psi_{-1} ({\bf r})
        \end{array}
 \right)
        = \sqrt{n(r)}
          \left( \begin{array}{c}
                    \cos^2{\beta\over 2} \\
                    \sqrt 2e^{i\phi}\sin{\beta\over 2}\cos{\beta\over 2} \\
                    e^{2i\phi}\sin^2{\beta\over 2}
                 \end{array}
          \right)
 \label{eq:MH}
\end{eqnarray}
where the bending angle $\beta(r)$ runs over $0\leq\beta(r)
\leq\pi$ and $\phi$ signifies the polar angle in polar coordinates. 
The spin direction is denoted by
the $l$-vector and is given as $\mbox{\boldmath $l$}(r)={\hat z}\cos\beta+
\sin\beta (\cos\phi{\hat x}+\sin\phi{\hat y})$ where $\beta$ varies from
$\beta(0)=0$ to $\beta(R)=\frac{\pi}{2}\; (=\pi)$ for MH (Anderson-Toulouse (AT)) ($R$ is
the outer boundary of the cloud). Thus the spin moment is flared out to the radial
direction and at the circumference it points outward for MH and downwards for AT
(for schematic $l$-vector structure, see Fig.18 in Ref.\cite{salomaa} ).

%===================== <lz> ======================
\begin{figure}[b]
\includegraphics[width=8cm]{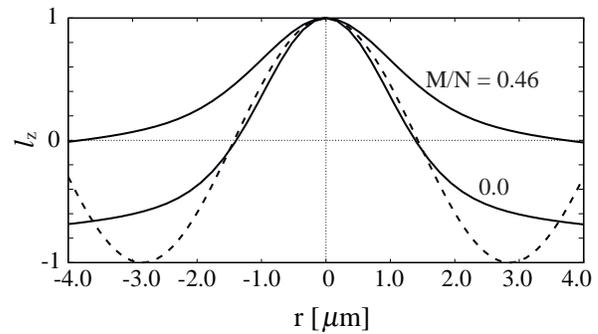}
\caption{\label{fig:lz_M}
Spatial dependence of the $l_z$-component along the radial direction 
at M/N=0, 0.46, and $\Omega=0.37$.
The dashed line shows cos$\beta(r)$ with the bending angle
$\beta(r)=\pi r/R$ ($R=2.85\mu m$)
for the AT vortex.
}
\end{figure}
%===================== <lz> ======================/

In Fig.\ref{fig:lz_M} we show the spatial dependence of the $l_z$-component 
along the radial direction, namely, the spatial dependence of the bending
angle $\beta(r)$ for the MH vortex. 
As the magnetization $M$ decreases,
the local magnetization in the condensate surface 
changes from positive to negative passing through zero.
It means that the $l$-vector in this vortex flares out radially to orient almost horizontally
$\beta(r=R)={\pi\over 2}$ for $M/N \sim 0.5$ and to point downward for 
$\beta(r=R)=\pi$ for $M/N \sim 0$. 
The former (latter) corresponds literally to 
the Mermin-Ho (Anderson-Toulouse) vortex. 
This is simply because as $M$ decreases,
the spin-down component $\psi_{-1}$ with $w=2$ increases in
the outer region. Thus we can control these MH and AT vortices by merely
changing the total magnetization. 

As pointed out in the previous paper\cite{mizushima},
however, the situation is completely different from the case of superfluid $^3$He-A
where the stability of the MH vortex is due to the constraint 
that the $l$-vector be perpendicular to the vessel wall\cite{salomaa}.
These vortex configurations in ferromagnetic BEC
are created naturally under the condition of a given total number and magnetization,
both of which are well controlled in a harmonic trap experiment.

In comparison with the $\langle 0, 1, 2 \rangle$ vortex, 
the spin textures of other vortex configurations, 
such as the $\langle 1, \times, 0 \rangle$ and $\langle 1, 0, -1 \rangle$ vortices, have
a different nature\cite{isoshima1}. 
In $\langle 1, \times, 0 \rangle$ vortex, the spin moment  
is suddenly reversed near the vortex core of $\psi_1$ component 
because of the absence of $\psi_0$ component.
In $\langle 1, 0, -1 \rangle$ vortex it can vary continuously around the vortex core. 
In this configuration, however, 
since the condensate at the center of the trap consists only of the polar state\cite{ho}, 
the spin texture has a singularity.  
Thus only $\langle 0,1,2 \rangle$ vortex can have a non-singular and continuous 
spin texture under slow rotation. 
 
%===================== <012gs> ======================
\begin{figure}[t]
  \includegraphics[width=8.5cm]{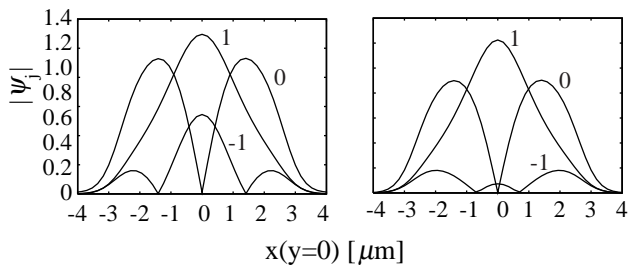}
\caption{\label{fig:012gs}
Density profiles of the $\langle 0, 1, 2 \rangle$ vortex
in an antiferromagnetic interaction of $g_s/g_n = 0.005$ (on the left)
and the non-magnetic interaction of $g_s/g_n = 0$ (on the right) situation.
}
\end{figure}
%===================== <012gs> ======================/

It is easy to calculate the total angular momentum $L_z$ of the axisymmetric vortices;
by using the total particle number and the total magnetization,
it is simply written as
\begin{eqnarray}
 \frac{L_z}{\hbar N} = w_0 + w' \frac{M}{N},
 \label{eq:lz}
\end{eqnarray}
where the total magnetization is written as 
$M=\int d{\bf r} n({\bf r}) \hat{l}_z$ 
and we have introduced $w_j - w_0 = j w'$.
Thus the spin textures with a net spin polarization carry 
the angular momentum, i.e. the superflow.
For $\langle 0, 1, 2 \rangle$ vortex, $\frac{L_z}{\hbar N} = 1 - \frac{M}{N}$. 
This simple formula has the following physical meaning.
(i)At $M\!=\!N$, $L_z\!=\!0$ because $\psi_1$ has no winding.
(ii)At ${M\over N}\!=\!{1\over 2}$, ${L_z\over \hbar N}$ is exactly equal to $\hbar/2$,
corresponding to the MH vortex.
 
In the $\langle 0, 1, 2 \rangle$ state with the higher winding,
the non-winding component $\psi_1$ works as a ``pinning potential'' 
for the remaining $\psi_0$ and $\psi_{-1}$, thereby making the state 
stable in the lower rotation drive.
In particular, $\psi_{-1}$ with $w_{-1}\!=\!2$ 
is stabilized by the presence of the $\psi_1$ due to the ferromagnetic interaction. 
For a very small antiferromagnetic interaction ($g_s=0.005g_n$) 
and non-magnetic case ($g_s/g_n \!=\! 0$), 
the vortex with the $w_{-1}\!=\!2$ becomes unstable 
and splits into a couple of $w_{-1}\!=\!1$ vortices (see Fig.\ref{fig:012gs}). 
This configuration is equivalent to the vortex found by Yip (phase IV in Ref.\cite{yip}) 
and is always unstable for the large $g_s/g_n$ ($>0$).

%**************** Non-Axisymmetric vortex **************
\subsection{Non-Axisymmetric vortex}

%===================== <111gs> ======================
\begin{figure}[t]
  \includegraphics[width=8.5cm]{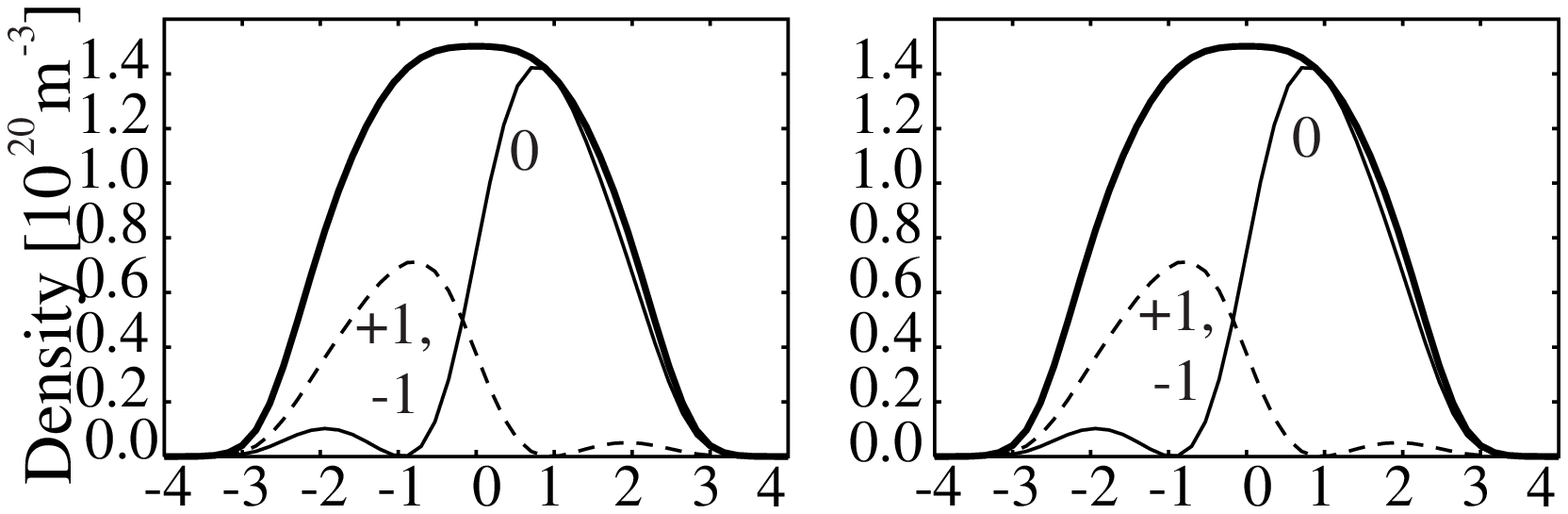} \\
  (a) \\
  \includegraphics[width=8.5cm]{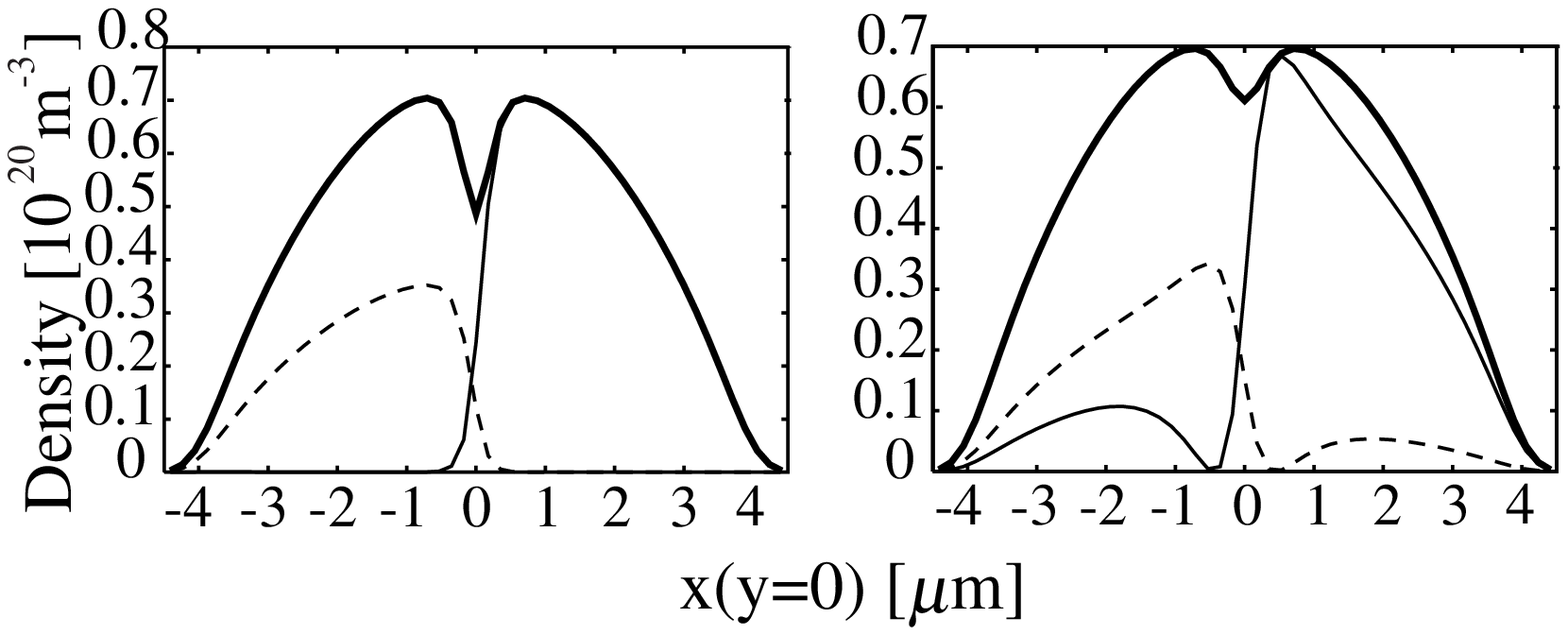} \\
  (b) \\
\caption{\label{fig:111gs}
Density profiles in the $\langle 1,1,1 \rangle$ split-(I) (on the left)
and split-(II) vortex (on the right) at $\Omega=0.35$ and $M/N = 0$:
(a) Non-magnetic interaction case ($g_s/g_n = 0.0$)
and (b) $g_s=0.4g_n$.
The bold line show the total density $n({\bf r})$
and the thin and dashed lines present
the density profiles of the internal structures, respectively.
}
\end{figure}
%===================== <111gs> ======================/

To investigate the possibility of non-axisymmetric vortices, 
let us first recapitulate the axisymmetric $\langle 1,1,1 \rangle$ vortex. 
In the axisymmetric case, the total density of the  
$\langle 1, 1, 1 \rangle$ vortex is equivalent to the one in scalar BEC,
 having the singularity at the center of the trap
 where the potential energy is minimum.
The axisymmetric singular $\langle 1, 1, 1 \rangle$ vortex 
is always unstable even in the higher rotation\cite{isoshima2}.
However, 
by displacing the vortex cores of each component from the center of the trap,
the $\langle 1, 1, 1 \rangle$ vortex can be stable as a non-axisymmetric non-singular type
in the lower rotation frequency.
This works favorably to gain more condensation energy at the center of the trap.
A similar situation is seen near the $\Omega_{c2}$ in the homogeneous system,
where the singular vortex lattice, called the Abrikosov lattices, 
is never favored in the entire antiferromagnetic region\cite{kita}.

For the antiferromagnetic case, 
$\psi_1$ and $\psi_{-1}$ overlap 
to minimize the spin-dependent energy $E_s ({\bf r})$.
In Fig.\ref{fig:111gs}, 
we show the density profiles and $g_s$-dependence of two different 
$\langle 1, 1, 1 \rangle$ vortices.
In the non-magnetic limit ($g_s/g_n\!=\!0$), 
the two states are completely equivalent.
As the spin interaction $g_s/g_n$ increases,
striking differences grow between the two states.
For $g_s/g_n\! \sim \!0.4$, 
the vortex cores of the state presented in the left of Fig.\ref{fig:111gs} (b)
(the split-(I) state) collapse and the amplitude of each order parameter 
cannot recover near the vortex cores, 
i.e. this state behaves like the phase separation in $x$-$y$ plane.
In contrast, the state shown in the right of Fig.\ref{fig:111gs} (b)
(the split-(II) state) forms the regular cores.
As a result, the split-(II) state is energetically favorable over the 
split-(I) state for the antiferromagnetic situation.
In very small spin interaction range ($g_s\sim0.02g_n$),
which is a realistic parameter, however,
the split-(I) state can be favored over the split-(II),
though the energies of two states are very close to each other.

%===================== <111tri> ======================
\begin{figure}[t]
    (a) \includegraphics[width=7.5cm]{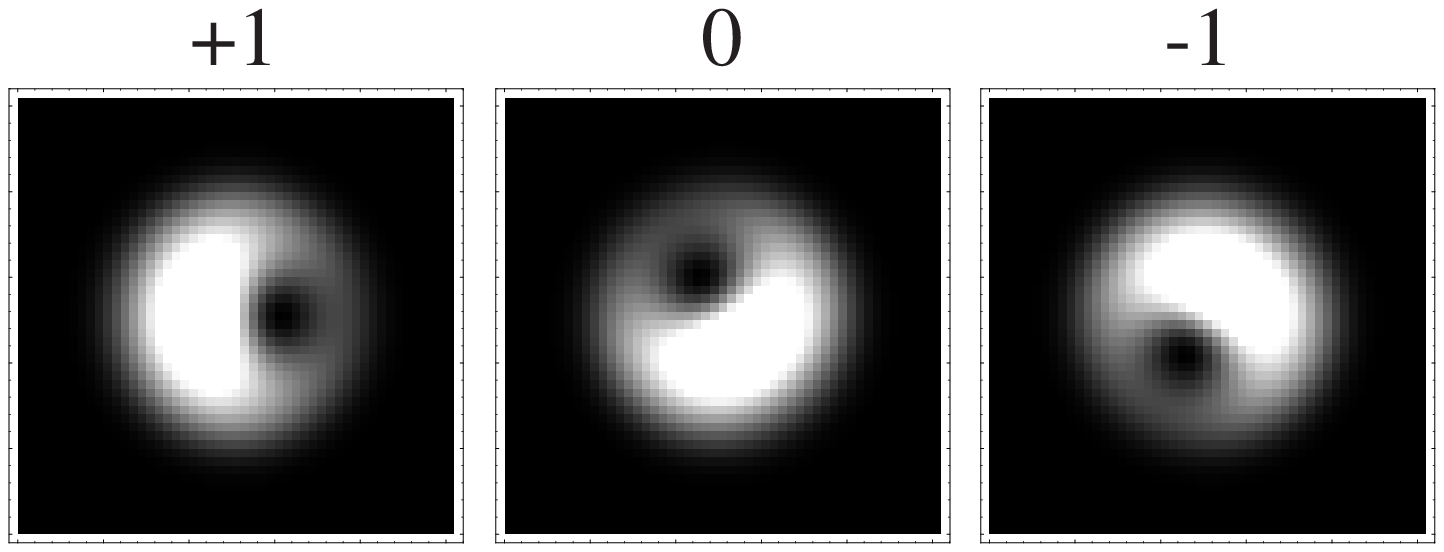} \\
    (b) \includegraphics[width=7.5cm]{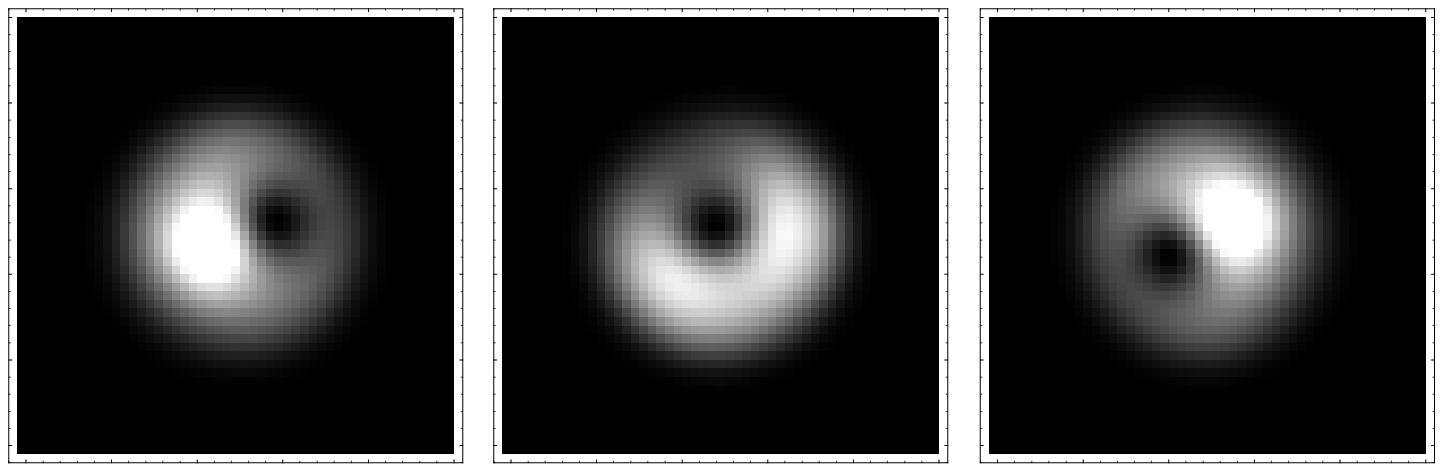} \\
    (c) \includegraphics[width=7.5cm]{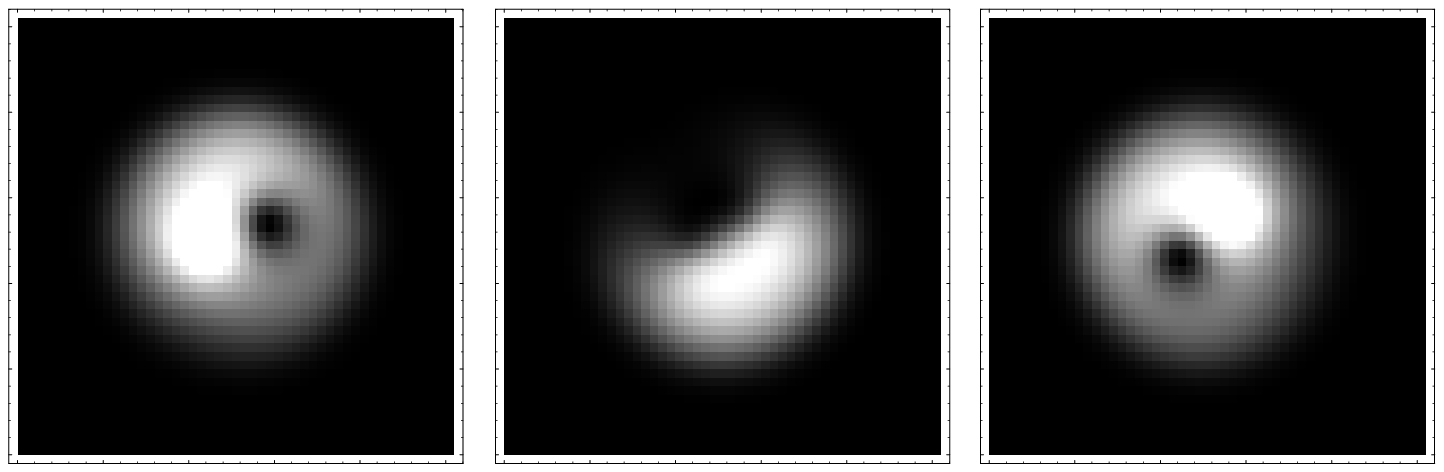} \\
\caption{\label{fig:111tri}
Density profiles of each component 
$|\psi_1({\bf r})|^2$, $|\psi_0({\bf r})|^2$, and $|\psi_{-1}({\bf r})|^2$
in $\langle 1,1,1 \rangle$ triangle vortex at $\Omega=0.35$ and $M/N = 0$:
(a) Non-magnetic interaction case ($g_s/g_n = 0.0$), 
(b) $g_s=0.02g_n$, and (c) $g_s=0.2g_n$.
The total density profile almost agrees with that in the vortex-free state, i.e. 
the non-singular type.
}
\end{figure}
%===================== <111tri> ======================/

A third vortex configuration is displayed in Fig.\ref{fig:111tri}
where vortex cores of each component are displaced to form a triangle.
In the non-magnetic limit, the three components are completely equivalent
and the three singularities form a regular triangle.
As the spin interaction $g_s$ increases, 
the $\psi_1$ and $\psi_{-1}$ components overlap locally 
because of the antiferromagnetic interaction
and this vortex configuration starts to deform from a regular triangle shape. 
For larger spin interactions ($g_s>0.4$), 
this vortex becomes unstable due to the overlap between $\psi_1$ and $\psi_{-1}$ components.

%===================== <111> ======================
\begin{figure}[t]
%\begin{center}
    \includegraphics[width=7cm]{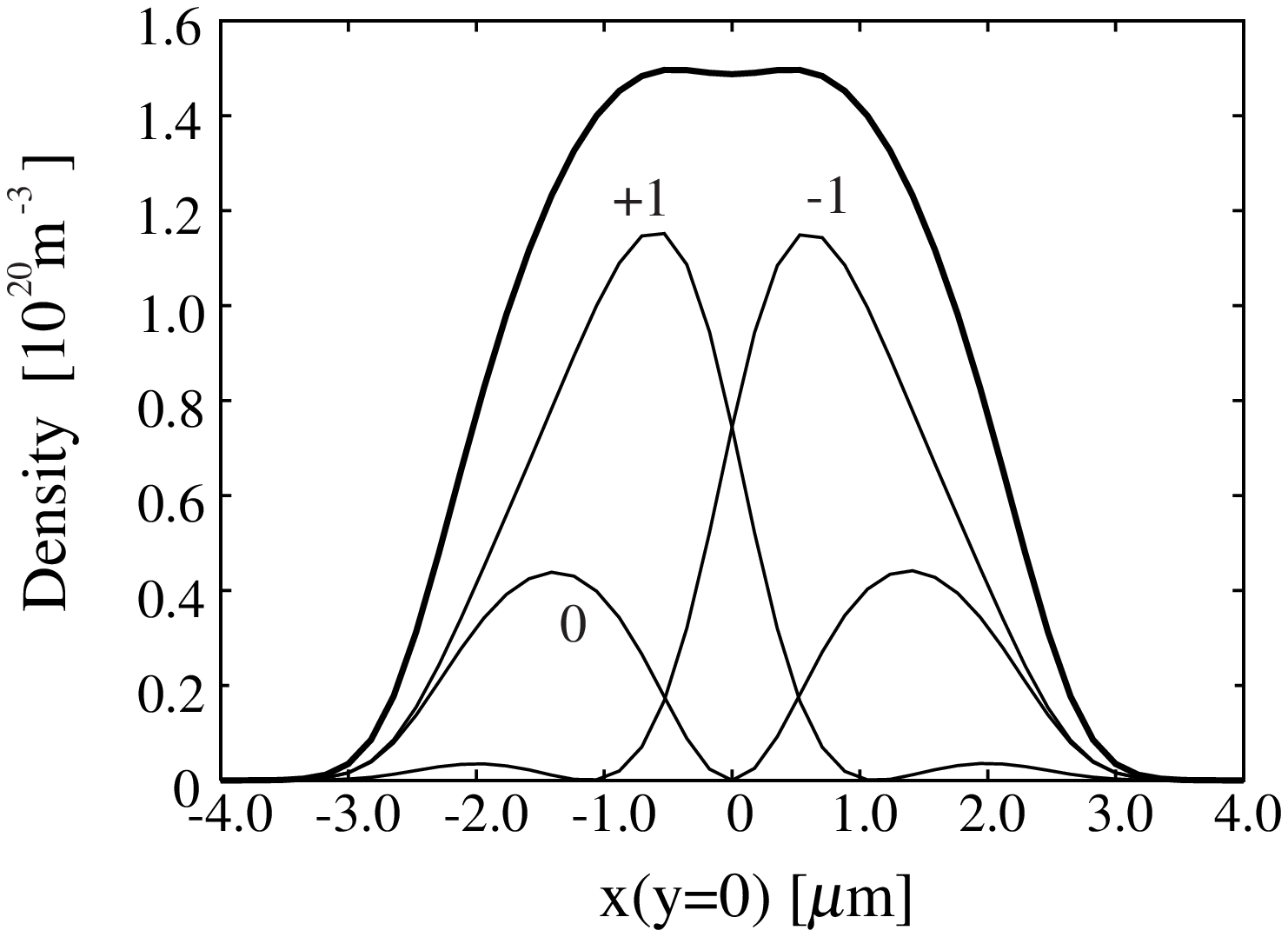} \\
    (a) \\
 \begin{tabular}{cc}
    \includegraphics[width=4.5cm]{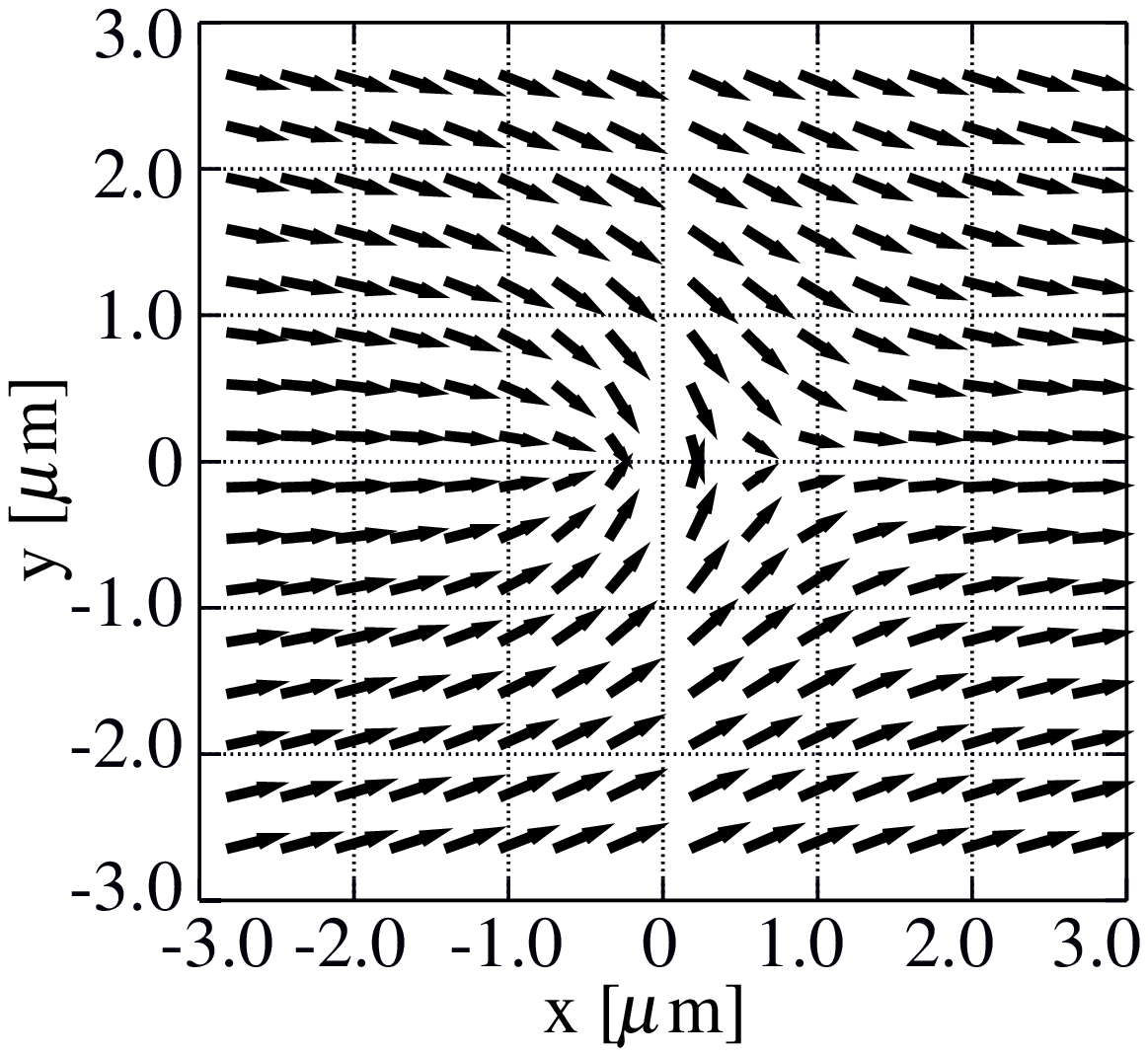}
    &
    \includegraphics[width=4cm]{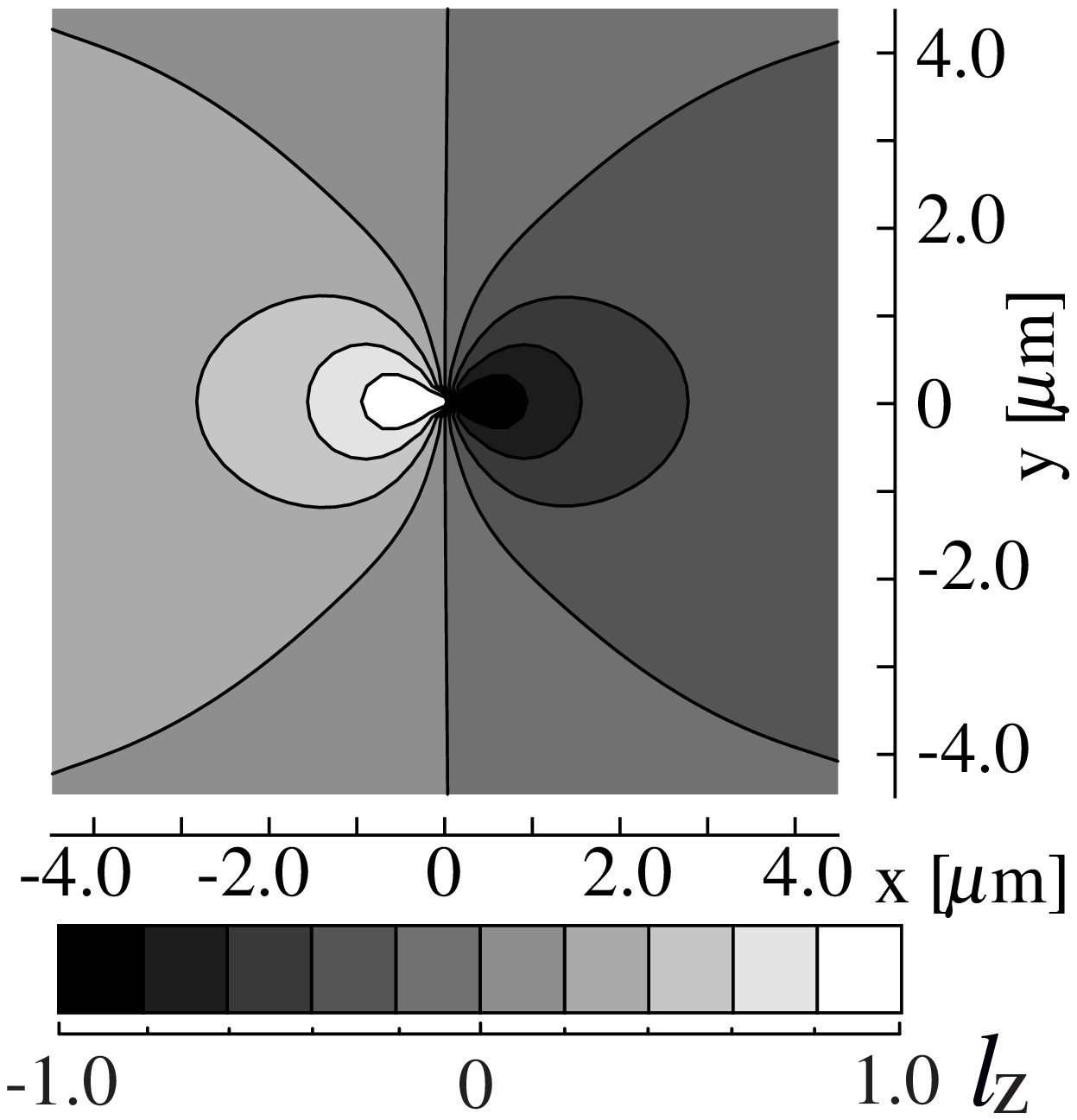}\\
    (b)
    &
    (c)
  \end{tabular}
%\end{center}
\caption{\label{fig:111dns}
(a)Density profile of the condensates,
(b)the $l_x$ and $l_y$
and (c)the density map of $l_z$
for the non-axis-symmetric $\langle 1,1,1\rangle$ vortex
in $\Omega=0.35$ and  M/N=0.0.
The bold line is the total density $n({\bf r})$
and the thin lines show the density of each component $|\psi_j|^2$.
}
\end{figure}
%===================== <111> ======================/

Figure \ref{fig:111dns} (a) shows the density profiles of 
a non-axisymmetric non-singular $\langle 1, 1, 1 \rangle$ vortex
for the ferromagnetic case.
Two singularities of $\psi_1$ and $\psi_{-1}$ are displaced from the center.
The $\psi_0$ component with the singularity at the center of the trap
plays the role to prevent the phase separation 
favored in the ferromagnetic spin interaction.
The spin texture in this state is displayed in Figs.\ref{fig:111dns} (b) and (c)
where the spin moments flip at the center of the trap.

In comparison with axisymmetric types, 
these non-axisymmetric vortices have the advantage 
that they can easily adjust themselves for a change in $\Omega$. 
As $\Omega$ increases, the two or three separate singularities adjust their mutual distance from the 
center and change the value of $L_z$ to gain the energy from the term $-\Omega L_z$. 
In this sense, the non-axisymmetric vortices are flexible against a change in $\Omega$
compared with the axisymmetric ones.

%###################### PHASE DIAGAM #############################
\section{phase diagram}

The phase diagrams are shown in a plane of the external rotation and 
the total magnetization
by comparing the energies of various vortex configurations:
%===================== Energy =====================================
\begin{eqnarray}
 E = \int d {\bf r} \left[ \sum_i E_i ({\bf r}) + E_s ({\bf r}) \right]
     - \Omega L_z.
\label{eq:ene}
\end{eqnarray}
%===================== Energy =====================================/ 
It is noted that the region $\Omega\!<\!0.38$ considered here 
corresponds to the single vortex region in the scalar BEC case\cite{isoshima3}.
Thus, the present single-vortex consideration may also be justified for $\Omega<0.38$.

%===================== Phase ======================
\begin{figure*}[t]
% \begin{center}
  (a) \includegraphics[width=8cm]{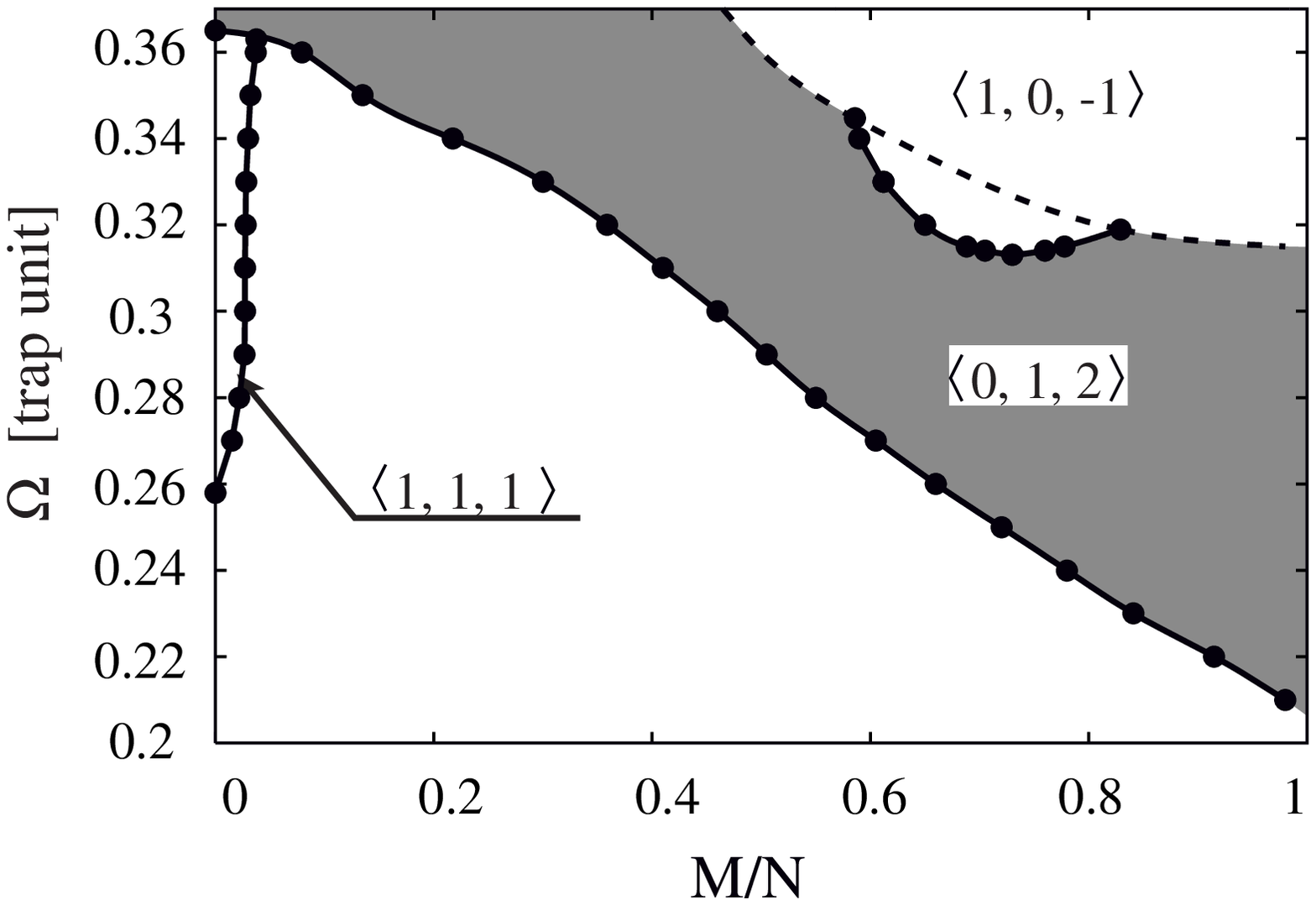} 
  (b) \includegraphics[width=8cm]{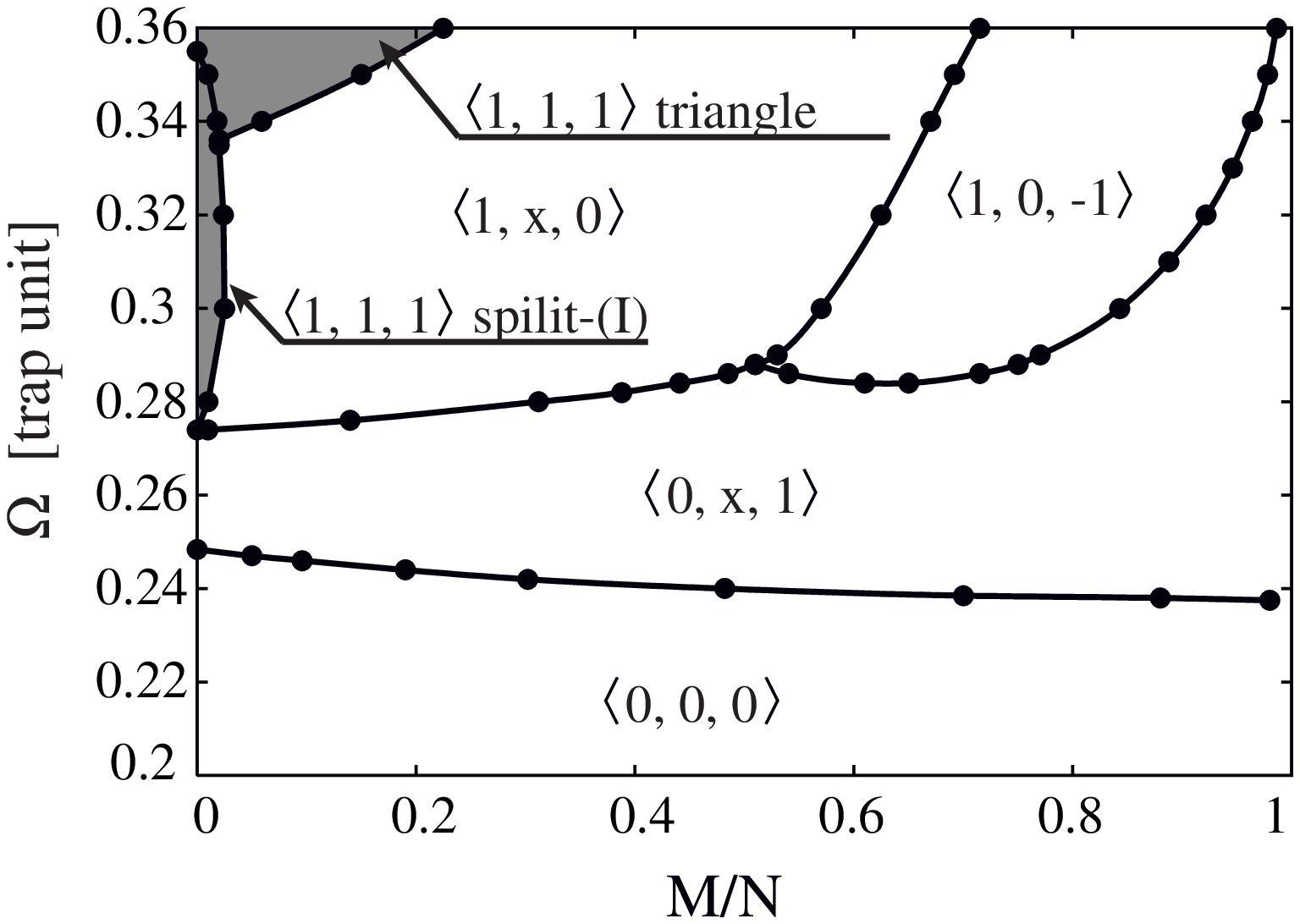}
% \end{center}
\caption{
   \label{fig:diagram}
Phase diagram for (a)the ferromagnetic state ($g_s=-0.02g_n$)
and (b)the antiferromagnetic case ($g_s=0.02g_n$).
The dashed line in (a) denotes the boundary where the lowest quasiparticle energy of
the $\langle0,1,2\rangle$ vortex becomes negative\cite{mizushima}.
}
\end{figure*}
%===================== Phase ======================/

The resulting phase diagram is displayed in Fig.\ref{fig:diagram} for
(a) the ferromagnetic case and (b) the antiferromagnetic case.
For the ferromagnetic case,  
a large area of the $\Omega-M$ plane
is occupied by the $\langle 0,1,2\rangle$  vortex, including MH and AT.
The non-axisymmetric  $\langle 1,1,1\rangle$  vortex
and the $\langle 1,0,-1\rangle$  vortex are stabilized 
near $M\!=\!0$ and $M\!=\!N$, respectively.
We find a large empty area
in the intermediate $M/N$ region where neither single-vortex nor 
vortex-free states are stabilized at all because the phase separation in
the ferromagnetic case prevents forming a uniform mixture
of the three components even when the circulation is absent in the
vortex-free state.

In the phase diagram for the antiferromagnetic case, in contrast, 
everywhere is occupied by a stable phase.
This result is consistent with Fig.2(a) of Ref.\cite{isoshima2} over a wide range, 
except for the presence of two non-axisymmetric types near $M \sim 0$, 
i.e. $\langle 1, 1, 1 \rangle$ split-(I) and triangle vortex.
The $\langle 1, 1, 1 \rangle$ triangle vortex is energetically indistinguishable from 
the phase-I vortex obtained by Yip\cite{yip}.  
The phase-IV vortex given in Fig,3 of Ref.\cite{yip} is found  unstable for 
$g_s\!=\! 0.02 g_n$ and does not appear in this phase diagram.

%===================== Angular Momentum ======================
\begin{figure}[b]
 \begin{center}
  (a) \includegraphics[width=7.5cm]{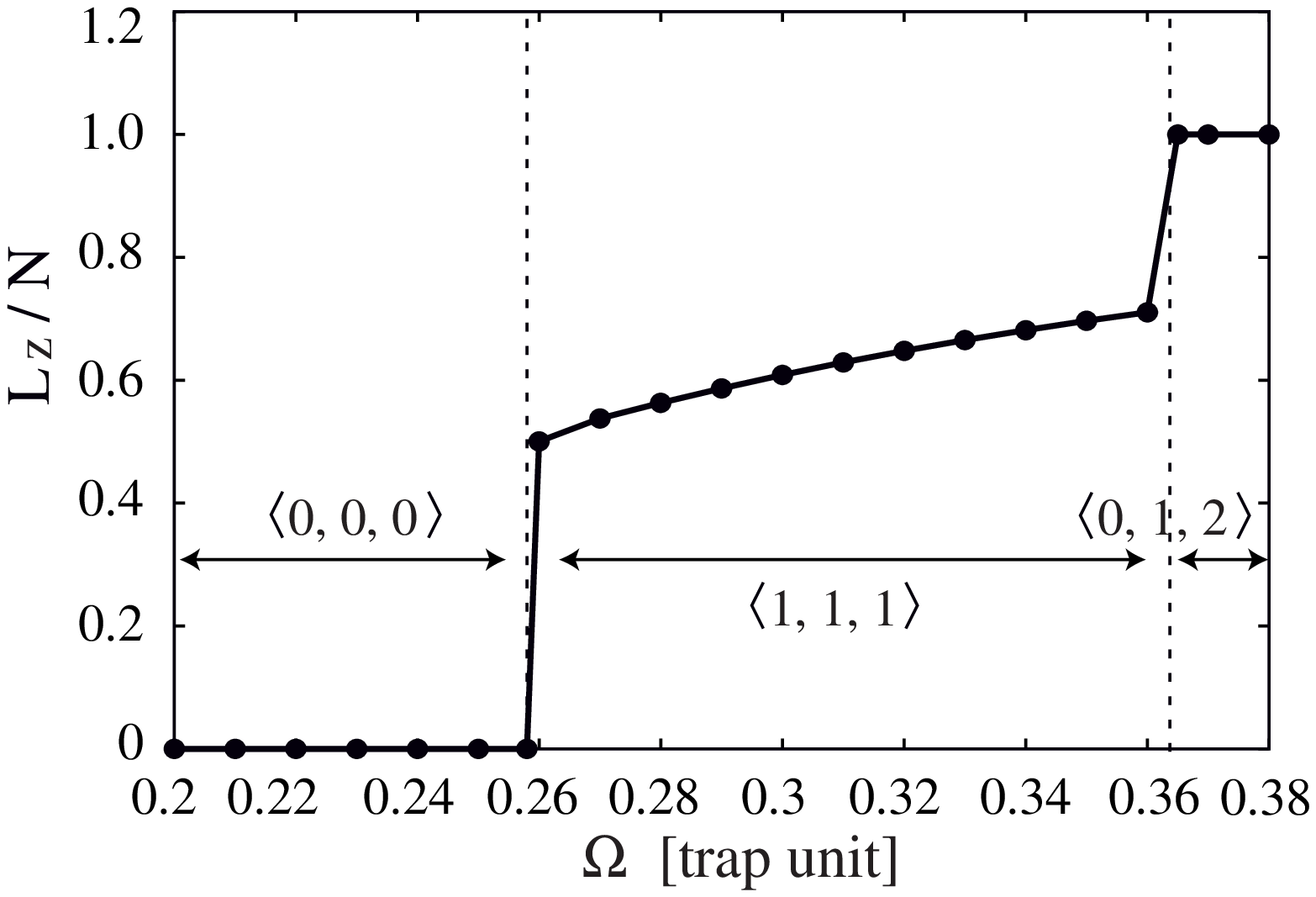} \\
  (b) \includegraphics[width=7.5cm]{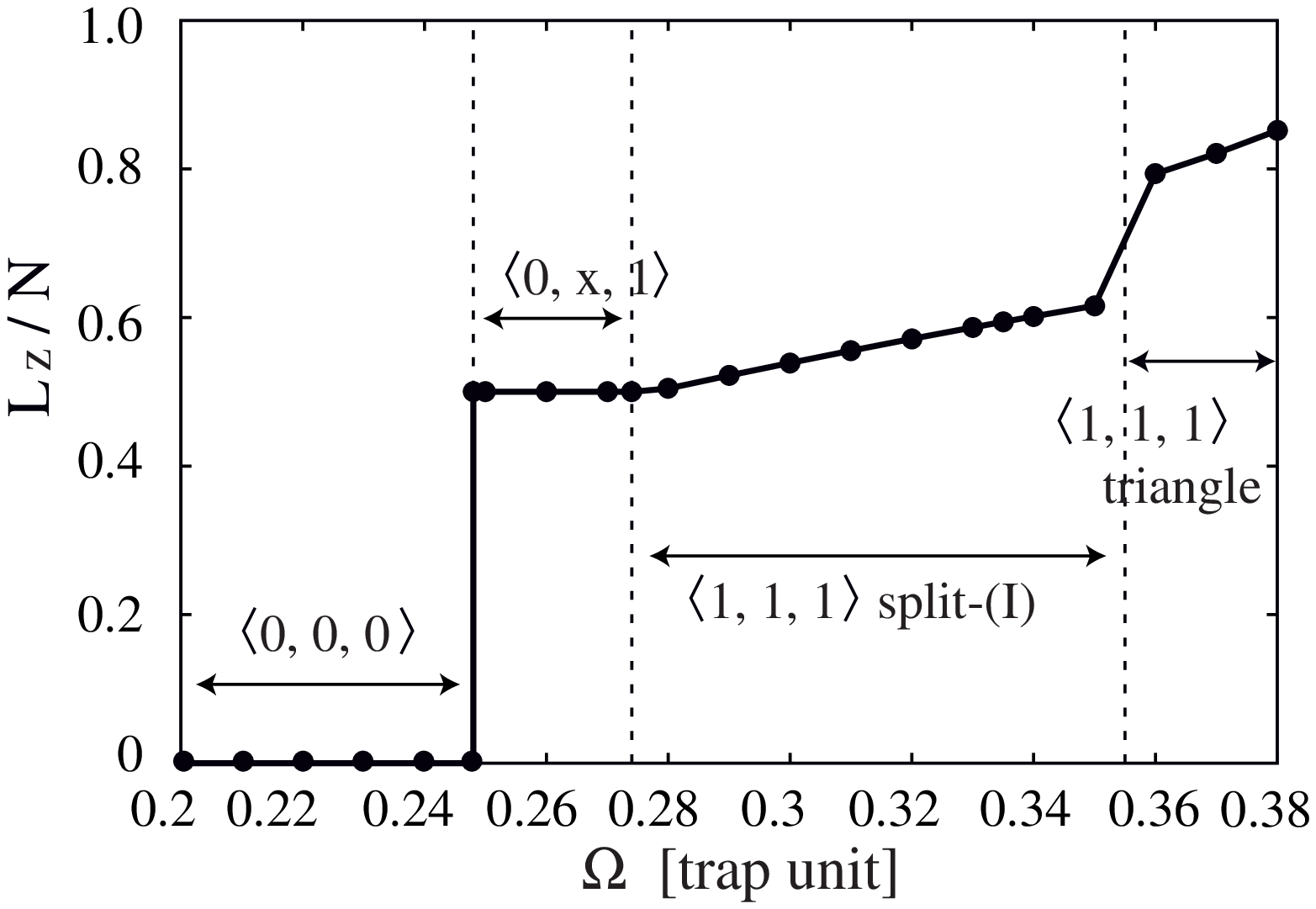}
 \end{center}
\caption{
   \label{fig:AFLz}
$\Omega$-dependence of the angular momentum $L_z/N$ at $M/N=0$:
(a) the ferromagnetic case ($g_s \!=\! -0.02g_n$) and
(b) the antiferromagnetic case ($g_s \!=\! 0.02g_n$).
}
\end{figure}
%\end{figure*}
%===================== Angular Momentum ======================/

It is found for both cases that the stabilities of the non-axisymmetric vortices 
are restricted in a narrow region. 
On the other hand, the axisymmetric vortices 
are stable in a large area. 
As discussed in Section III,
since finite density of a component at the vortex cores of the others 
supports its stability, the non-axisymmetric vortex
becomes unstable in increasing $M$
where the number of $\psi_1$ grows while the others shrink.

Figure \ref{fig:AFLz} shows the $\Omega$-dependence of the angular momentum at $M=0$
for the two cases.
For axisymmetric types, as shown in Eq.(\ref{eq:lz}), 
the angular momentum of the system $L_z$ is fixed for a given $M$. 
Thus there is the need of changing the winding combinations $\langle w_1,w_0,w_{-1} \rangle$
so as to increase $L_z$, which 
means that the axisymmetric vortices do not have the adaptability
for changes in $\Omega$.

%###################### Conclusion and discussion #############################
\section{Discussion}

The phase-separated state with $w\!=\!0$ is expected to be stable 
in a large empty region of Fig.7 (a).
In this state, 
$\psi_1$ and $\psi_{-1}$ components phase-separate along $z$-direction
due to the ferromagnetic interaction.
Namely, an arbitrary $x$-$y$ cross-section consists of only $\psi_1$ or $\psi_{-1}$
component, and the spin-polarized state with $M/N\!=\! \pm 1$ is piled up along the z-direction. Neglecting
the contribution from the boundary layer associated 
with the phase separation, we can estimate that the energy
of this phase-separated state with $w\!=\!0$ and $w\!=\!1$
is equal to the energy of the scalar BEC with $w\!=\!0$ and $w\!=\!1$, respectively.

\begin{figure}[t]
 \begin{center}
  (a) \includegraphics[width=7.5cm]{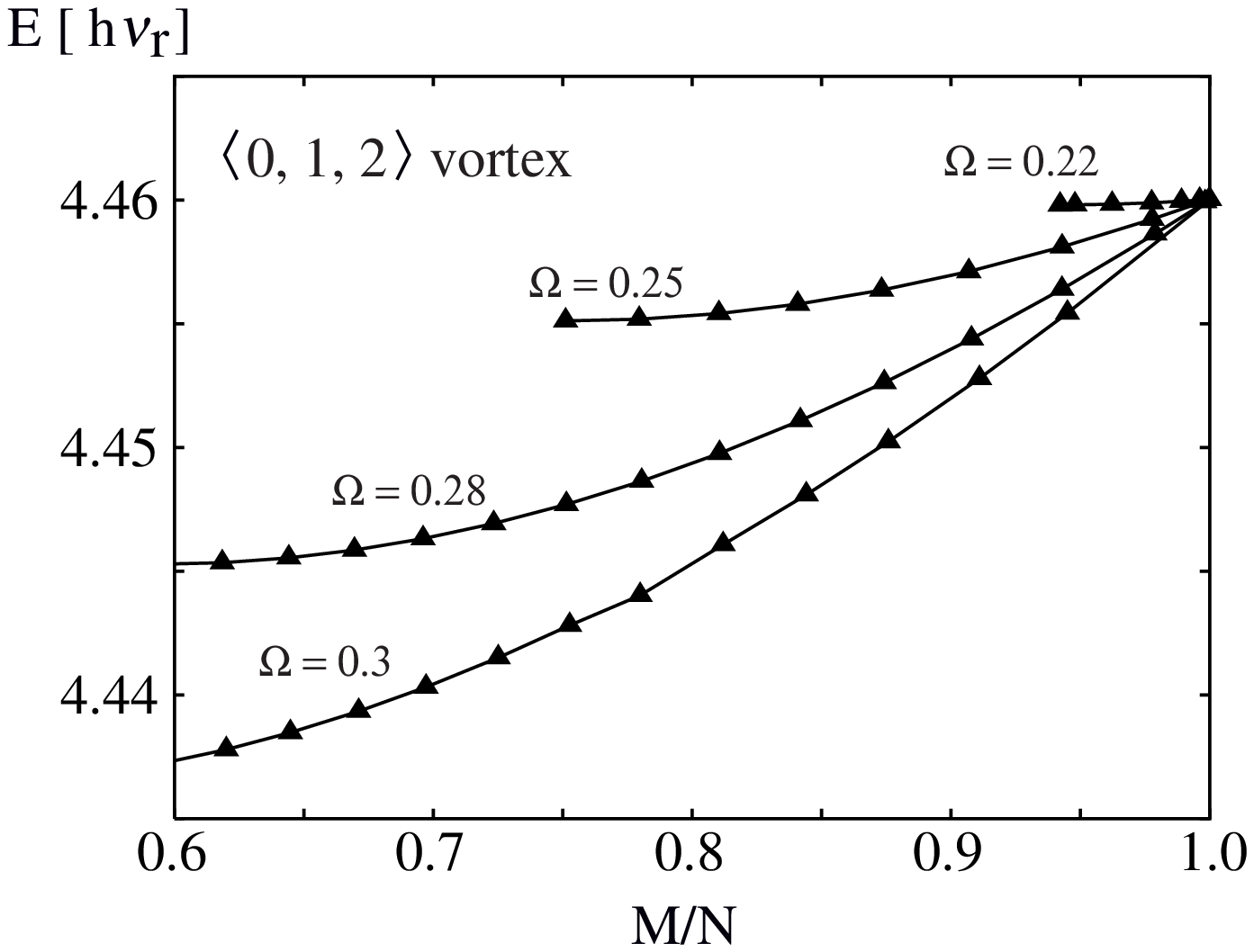}\\ 
  (b) \includegraphics[width=7.5cm]{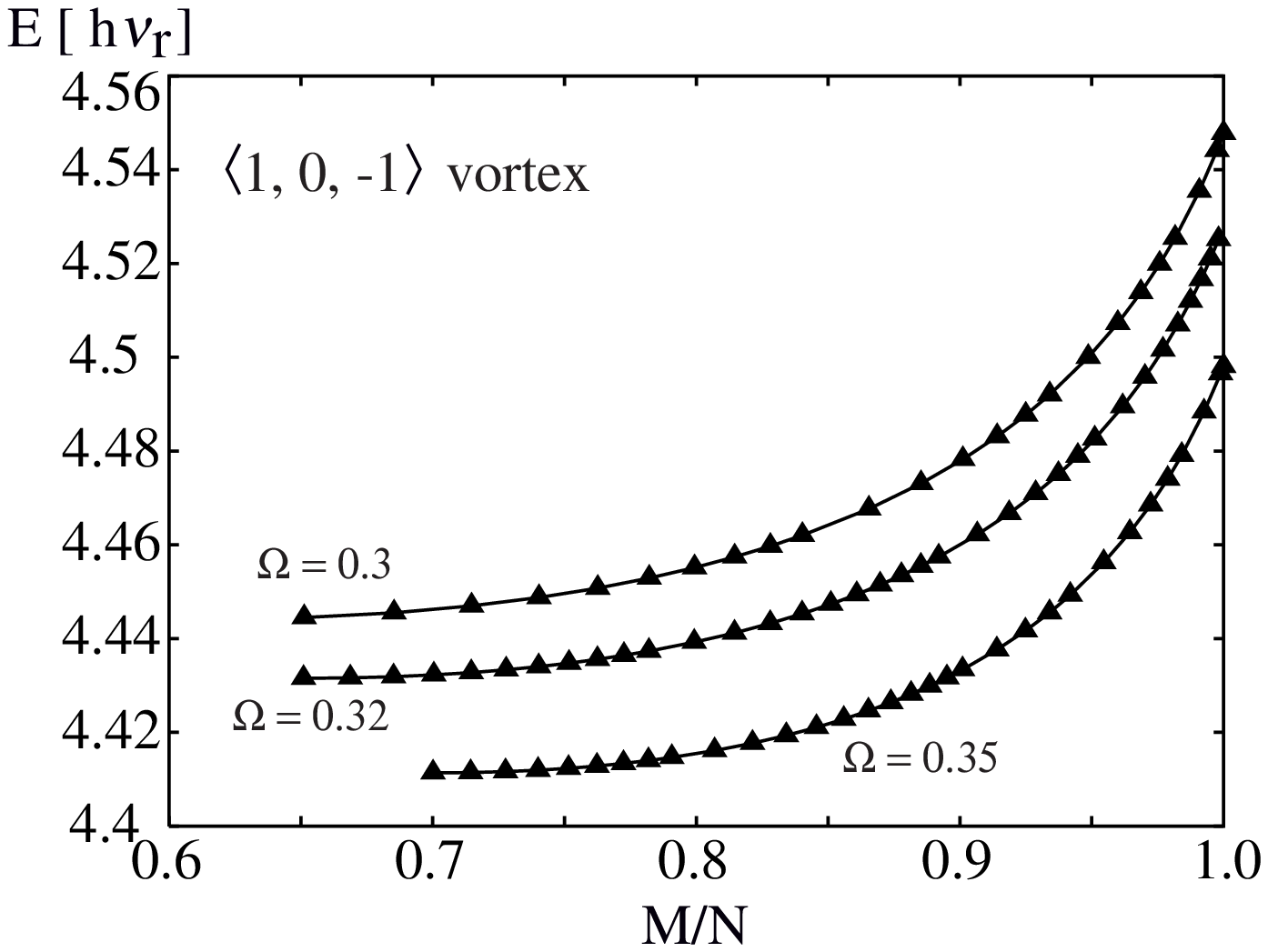}
 \end{center}
\caption{
   \label{fig:Ene_mag}
The $M$-dependence of the energy for the ferromagnetic case:
(a)the $\langle 0, 1, 2 \rangle$ and (b)the $\langle 1,0,-1 \rangle$
vortices.
The energies of the phase-separated states with $w\!=\!0$ and $w\!=\!1$
correspond to the energies of the $\langle 0,1,2 \rangle$ vortex at $M/N\!=\!1$ and 
$\langle 1,0,-1 \rangle$ vortex at $M/N\!=\!1$, respectively.
}
\end{figure}

We compare in Fig.\ref{fig:Ene_mag} (a)
the free energies of the $\langle 0,1,2 \rangle$ vortex state
and the phase-separated state.
As shown in  Fig.\ref{fig:Ene_mag} (a),
the $\langle 0,1,2 \rangle$ vortex with the three components 
is energetically favored over the phase-separated state, 
where the energy of the phase-separated state is given by the energy of 
the $\langle 0,1,2 \rangle$ vortex at $M/N\!=\!1$, i.e. the scalar BEC with $w\!=\!0$.
Thus, the composite state of the three components, such as the MH vortex,
becomes ``locally'' stable under a rotation drive 
while the composite state may have ``global'' stability.
It is possible to perform a similar discussion for the phase-separated state
in higher rotation.
The result of Fig.\ref{fig:Ene_mag} (b) shows that the $\langle 1,0,-1 \rangle$ vortex
is favored over the phase-separated state with the winding,
whose total density corresponds to the conventional singular vortex in the scalar BEC.

%===================== Ferro Phase ======================
%===================== Ferro Phase ======================/

%#######################
\section{Conclusion}

We have presented possible vortex structures and 
the vortex phase diagram in the plane of 
external rotation $\Omega$ and the total magnetization $M/N$ 
for the both cases of antiferromagnetic ($g_s = 0.02g_n$) and 
ferromagnetic ($g_s=-0.02g_n$) interaction.  
We have investigated the thermodynamic stability of the possible vortex configurations
by solving the extended Gross-Pitaevskii equation for the spinor BEC with $F=1$. 

For the ferromagnetic case, 
the stability of the continuous vortex, 
called the Mermin-Ho and Anderson-Toulouse vortex, 
is demonstrated, 
but these topological structures are found never stable under no rotation drive. 
Furthermore, 
these vortices can exist in the intermediate process (see Fig.5 in the Ref.\cite{isoshima4}) 
proposed by Isoshima {\it et al}.\cite{isoshima4, nakahara},
i.e. it may be created by making use of spin texture.

We have also discovered a couple of new non-axisymmetric vortices
besides the two vortices found by Yip\cite{yip} for the antiferromagnetic case.
The conventional singular vortex is found to be never favored in spinor BEC for both cases.
It means that the total density profile is always non-singular and continuous. 
Therefore, the experimental procedure 
to image the magnetic patterns for each vortex configuration
will be required as a special technique to identify these vortices.

\acknowledgements
The authors thank T. Ohmi and T. Isoshima for useful discussions.
One of the authors (TM) would like to acknowledges 
the financial support of Japan Society for the Promotion
of Science (JSPS) for Young Scientists.

%%%% references %%%%%%%%%%%%%%%%%%%%%%%%%%%%%%%
%%%%%%%%
%\begin{references}

\end{document}